\documentclass[journal,10pt]{IEEEtran}

\usepackage[cmex10]{amsmath}
\usepackage{mathrsfs}
\usepackage{mathtools}
\usepackage{xcolor}

\usepackage{colortbl}
\usepackage[dvipsnames]{xcolor}
\usepackage{balance}
\usepackage{amsmath}
\usepackage{amssymb}
\usepackage{stfloats}
\usepackage{cite}
\usepackage{graphicx}
\usepackage{subfigure}
\usepackage{bm}
\usepackage{pifont}
\usepackage{algorithmic}
\usepackage[procnumbered,ruled]{algorithm2e}
\usepackage[normalem]{ulem}
\begin{document}

\title{Spatiotemporal 2-D Polar Codes over Non-Uniform MIMO Channels: A Reliability-Aware Construction Approach}

\author{Yaqi~Li,~\IEEEmembership{Student Member,~IEEE,}
	~Shuohan~Zhang,~\IEEEmembership{Student Member,~IEEE,}
	~Xiaohu~You,~\IEEEmembership{Fellow,~IEEE,}
	~and~Jiamin~Li,~\IEEEmembership{Member,~IEEE}
%	~and~Chen~Ji,~\IEEEmembership{Member,~IEEE,}

	% <-this % stops a space

	\thanks{This work was supported in part by Major Science and Technology Project of Jiangsu Province under Grant BG2024002, by the National Natural Science Foundation of China (NSFC) under Grants 62331009, by the Fundamental Research Funds for the Central Universities under Grant 2242022k60006, and by the Major Key Project of PCL (PCL2021A01-2). \emph{(Corresponding Author: Xiaohu You)}}
	
	\thanks{Y. Li, S. Zhang, X. You and J. Li are with the National Mobile Communications Research Laboratory, Southeast University, Nanjing 210096, China (e-mail:{230258100, 213223443, xhyu, jiaminli}@seu.edu.cn). X. You and J. Li are also with Purple Mountain Laboratories,
		Nanjing, 211111, China.}
	
%	\thanks{C. Ji is with School of Information Science and Technology, Nantong University, Nantong, 226019, China. (email: gwidjin@ntu.edu.cn)}
	
}

\maketitle

\begin{abstract}

With the increasing demand for ultra-reliable and low-latency communication (URLLC), spatiotemporal two-dimensional (2-D) channel coding has received growing interest. By leveraging the spatial degrees of freedom in massive multiple-input multiple-output (MIMO) systems, it shortens the time-domain blocklength, thereby reducing latency and enhancing reliability. However, existing spatiotemporal coding schemes typically assume uniform reliability across spatial streams. This assumption does not hold in practical MIMO channels, where the underlying propagation environment generally leads to unequal spatial-eigenmode gains and reliabilities, making the conventional Gaussian-approximation-based construction for 2-D polar codes less effective. This paper investigates spatiotemporal 2-D polar coding over non-uniform MIMO channels, where the spatial domain exhibits inherently heterogeneous signal-to-noise ratios (SNRs). We propose a reciprocal channel approximation (RCA)-based reliability-aware 2-D polar coding framework that accurately characterizes such heterogeneous SNRs without relying on log-likelihood-ratio distribution assumptions. Simulation results demonstrate that the proposed RCA-based spatiotemporal 2-D polar coding scheme achieves clear performance gains and strong robustness, confirming its effectiveness in jointly exploiting temporal and spatial polarization for URLLC in practical MIMO systems.

% However, existing spatiotemporal coding schemes typically assume uniform reliability across spatial streams, which does not hold in practical MIMO channels where different spatial eigenmodes generally exhibit unequal gains and reliabilities, and the standard Gaussian-approximation-based construction for 2-D polar codes becomes less effective in non-uniform MIMO channels.
	
\end{abstract}


\IEEEpeerreviewmaketitle

\section{Introduction}\label{sec1}

\IEEEPARstart{D}{riven} by the stringent reliability and latency requirements of ultra-reliable and low-latency communication (URLLC) in the fifth
generation (5G) and beyond systems, short blocklength coding has become indispensable for reducing latency \cite{saad2019vision,schulz2017latency}. However, the resulting blocklength reduction leads to the channel-capacity collapse effect in conventional time-domain coding \cite{you20236g}. To overcome this limitation, spatiotemporal two-dimensional (2-D) channel coding leverages the abundant spatial degrees of freedom (DoF) in massive MIMO systems to trade spatial DoFs for temporal DoFs, thereby enabling flexible reliability-latency tradeoffs and mitigating the performance degradation caused by short blocklengths \cite{10255714,Ye2025ExplicitFBLMIMO,you2022spatiotemporal}. To further approach the MIMO channel capacity, \cite{li2025performanceanalysisspatiotemporal2d} introduced spatiotemporal 2-D polar coding frameworks, where the inherent polarization structure is extended to incorporate spatial DoF, thereby strengthening polarization across both dimensions. These schemes provide capacity-approaching and latency-efficient channel coding solutions for massive MIMO systems. 

Existing studies on spatiotemporal 2-D channel coding have mainly focused on exploiting spatial DoF to alleviate the latency--reliability tension in short-blocklength transmission. However, they typically assume uniform reliability across spatial streams, which limits their applicability to practical MIMO systems. In realistic MIMO channels, the spatial streams generally exhibit heterogeneous channel qualities due to the unequal gains and corresponding reliability levels induced by the underlying propagation environment\cite{2000Capacity,tse2005fundamentals,1192168}. This heterogeneity is widely encountered in practical propagation scenarios and can become more pronounced as the antenna dimension increases. Such structural non-uniformity has a profound impact on the performance of coded MIMO transmission. Consequently, conventional coding constructions designed for uniform channels are inherently suboptimal, since they fail to exploit the reliability disparity across spatial streams. Coding schemes that explicitly account for this heterogeneity are therefore better suited to practical MIMO system design.

{Moreover, Gaussian-approximation (GA)-based density evolution has been widely adopted for polar code construction due to its low complexity and practical effectiveness\cite{910580,6279525}. However, since GA relies on the assumption of symmetric Gaussian log-likelihood ratios (LLRs), its accuracy may degrade in long-blocklength or low-signal-to-noise ratios (SNR) regimes\cite{9220133}. This limitation becomes more evident in practical MIMO channels, where heterogeneous per-stream SNRs can cause the LLRs to deviate from Gaussianity, especially for low-SNR streams. As the blocklength and the number of spatial streams increase, the resulting mismatch may become more pronounced, thereby limiting the effective exploitation of spatial polarization. To address this issue, the reciprocal channel approximation (RCA) was originally proposed by Chung \cite{2000On} and later validated by Ochiai \textit{et al.} \cite{Ochiai2023New} for 1-D polar codes. RCA avoids explicit LLR distribution assumptions and instead tracks reliability evolution through mutual-information reciprocity. As a result, it remains accurate over a wide SNR range and is therefore better suited to the heterogeneous reliability conditions encountered in non-uniform MIMO channels, enabling more effective spatiotemporal 2-D polar code design under practical MIMO operating conditions.

Motivated by these observations, this work develops a reliability-aware spatiotemporal 2-D polar coding framework for non-uniform MIMO channels. Unlike existing designs developed under uniform spatial-channel conditions, the proposed framework explicitly captures the non-uniform SNR distribution across spatial streams and incorporates RCA-based reliability evolution into the 2-D polarization process. This enables more accurate reliability evaluation and frozen-bit selection under realistic MIMO operating conditions. The main contributions of this paper are summarized as follows.

\begin{itemize}
	\item We formulate and study a new practical coding scenario, namely spatiotemporal 2-D polar coding over non-uniform MIMO channels with heterogeneous spatial reliabilities. By explicitly incorporating unequal spatial SNRs into the coding design, the proposed framework better matches realistic MIMO propagation conditions than conventional uniform-channel-based constructions.
	\item We develop an RCA-based reliability-aware construction method for the 2-D polarization process. By more accurately tracking the heterogeneous reliability evolution across spatial streams, the proposed method enables improved frozen-bit selection and more effective exploitation of spatial polarization.
	\item Simulation results demonstrate that the proposed scheme achieves clear performance gains and strong robustness under practical MIMO conditions, including imperfect channel state information (CSI), confirming the effectiveness of jointly exploiting temporal and spatial polarization for reliable MIMO transmission.
\end{itemize}

\section{System Model}\label{sec2}

\subsection{MIMO Channel Model}
Consider an MIMO channel $\mathbf{H}$ with $S$ transmit antennas and $L$ receive antennas, operating over quasi-static flat fading channels, where the random fading coefficients remain constant over the duration of each codeword. The relationship between the channel input and output can be expressed as:
\begin{equation}
	{\mathbf{Y}=\mathbf{H}\mathbf{X}+\mathbf{Z}},
\end{equation}
where $\mathbf{X}\in\mathbb{C}^{S\times T}$ represents the signal transmitted over $T$ time samples, and each column of $\mathbf{X}$ is normalized to have unit norm. This normalization ensures that the total transmit power across the $S$ antennas is fixed, corresponding to uniform power allocation with an average per-antenna power of $1/S$. $\mathbf{Y}\in\mathbb{C}^{L\times T}$ is the corresponding received signal. The channel matrix is $\mathbf{H}\in\mathbb{C}^{L\times S}$ which contains random complex fading elements. Each element is an i.i.d. $\mathcal{CN}(0,1)$ complex Gaussian random variable, which remains constant over the $T$ time samples. $\mathbf{Z}\in\mathbb{C}^{L\times T}$ is the additive noise signal at the receiver, which is independent of $\mathbf{H}$ and each of its elements follows a $\mathcal{CN}(0,1)$ complex Gaussian distribution. Under the condition of fixed or slowly varying large-scale fading, the channel variation between adjacent coding blocks is small, and the channel can be reasonably assumed to be quasi-stationary over several consecutive blocks \cite{JUNG20111316}.

\subsection{Spatiotemporal 2-D polar coding}
In conventional 1-D time-domain polar coding over a binary input discrete memoryless channel, channel combining and splitting operations transform the $N=2^{n}$ independent uses of the original channel into a series of polarized channels with different reliabilities. The $K$ most reliable polarized channels are selected for transmitting information bits, and the remaining $N-K$ channels are used for transmitting frozen bits. Thus, an original sequence $u_1^N=(u_1,u_2,...,u_N)$ consisting of $K$ information bits and $N-K$ frozen bits is encoded into a codeword $x_1^N$ in the following manner.
\begin{equation}\label{eqn-1}
	x_1^N=u_1^NF_N,
\end{equation}
where $F_N=F^{\otimes n}$ is the generator matrix.

Conventional time-domain polar coding requires long code lengths to approach capacity, resulting in high latency unsuitable for 6G URLLC scenarios. To address this limitation, a novel spatiotemporal 2-D polar coding scheme is proposed to incorporate the DoFs of spatiotemporal transmission to the inherent structure of polar codes \cite{li2025performanceanalysisspatiotemporal2d}. The scheme extends the polarization process to the spatial-temporal domain by leveraging the large number of spatial DoFs introduced by massive MIMO. The generator matrices of spatial component and time-domain component are defined as $F_S$ and $F_T$, respectively. $S=2^s$ is the number of bit streams in spatial domain, $T=2^t$ is the blocklength in time domain, $N=S\times T=2^n$ represents the total coding length, and $n=s+t$. According to the rules of Kronecker product, $F^{\otimes n}=F^{\otimes (s+t)}=F^{\otimes s}\otimes F^{\otimes t}$, i.e., $F_N=F_S\otimes F_T$. Thus, the joint spatiotemporal 2-D polar encoding process can be described as a linear mapping that $$\mathrm{Enc}: \mathbb{F}_2^{S\times T} \to \mathbb{F}_2^{S\times T}: (\cdot)  \mapsto F_S (\cdot) F_T, $$ where $ \mathbb{F}_2$ is the binary field. Through this approach, spatial DoF can be more effectively exploited to achieve an effective trade-off between latency and reliability, enabling flexible allocation of time-domain and spatial-domain resources.

\section{Spatiotemporal Polar Coding Design over Non-Uniform MIMO Channels}\label{sec3}

Previous studies on spatiotemporal 2-D polar coding typically rely on the assumption that all spatial streams experience identical reliability. However, practical MIMO systems exhibit inherently non-uniform spatial-domain reliabilities, and applying coding designs based on the uniform-channel assumption can lead to suboptimal performance and even reliability degradation. To address this mismatch, we develop a spatiotemporal 2-D polar coding framework explicitly designed for non-uniform MIMO channels, together with its associated construction strategy.

\subsection{Parallel Channel Model and Spatial Non-Uniformity}
The original MIMO channel can be decomposed into a set of independent, memoryless subchannels using singular value decomposition (SVD). For an $L\times S$ MIMO channel matrix $\mathbf{H}$, its SVD can be expressed as
\begin{equation}
	\mathbf{H} = \mathbf{U}\boldsymbol{\Sigma}\mathbf{V}^H,
\end{equation}
where $\mathbf{U}\in\mathbb{C}^{L\times L}$ and $\mathbf{V}\in\mathbb{C}^{S\times S}$ are unitary matrices satisfying $\mathbf{U}\mathbf{U}^H=\mathbf{I}_L$ and $\mathbf{V}\mathbf{V}^H=\mathbf{I}_S$, respectively.  
The diagonal matrix $\boldsymbol{\Sigma}=\mathrm{diag}(\sqrt{\lambda_1},\sqrt{\lambda_2},\ldots,\sqrt{\lambda_m})$ contains the non-negative singular values of $\mathbf{H}$ in descending order. These singular values provide a natural characterization of the channel's spatial non-uniformity, where $m=\min\{S,L\}$ and $\lambda_k$ denotes the $k$-th eigenvalue of $\mathbf{H}\mathbf{H}^H$. 

By applying precoding and post-processing at the transmitter and receiver, respectively, the overall transmission process can be represented as
\begin{equation}
	\tilde{\mathbf{Y}} = \mathbf{U}^H\mathbf{Y} = \mathbf{U}^H(\mathbf{H}\mathbf{V}\tilde{\mathbf{X}} + \mathbf{Z}) = \boldsymbol{\Sigma}\tilde{\mathbf{X}} + \tilde{\mathbf{Z}},
\end{equation}
where $\tilde{\mathbf{X}}$ is the data symbol before precoding, $\mathbf{X}=\mathbf{V}\tilde{\mathbf{X}}$ denotes the precoded transmit signal, and $\tilde{\mathbf{Y}}=\mathbf{U}^H\mathbf{Y}$ denotes the received signal after linear combining. This SVD-based transmission model assumes transmitter-side CSI for spatial precoding, while the receiver uses the corresponding combining matrix for spatial-domain decomposition. The equivalent noise $\tilde{\mathbf{Z}}=\mathbf{U}^H\mathbf{Z}$ has the same statistical distribution as $\mathbf{Z}$ since $\mathbf{U}$ is unitary.

Therefore, the original MIMO channel $W$ can be equivalently represented as a set of $m$ parallel, independent AWGN subchannels with unequal channel gains:
\begin{equation}
	y_k = \sqrt{\lambda_k}\,x_k + n_k,\quad k=1,2,\ldots,m,
\end{equation}
which can be denoted as $\{W_k\,|\,k=1,2,\ldots,m\}$.  
Each subchannel $W_k$ is characterized by a distinct fading coefficient $\sqrt{\lambda_k}$, while the noise term $n_k$ remains circularly symmetric complex Gaussian with distribution $\mathcal{CN}(0,1)$. Figure \ref{fig1} illustrates the distribution of subchannel gains under different MIMO configurations. It can be observed that the singular values of the channel matrix $\mathbf{H}$ decay rapidly, resulting in pronounced differences in the reliability $\lambda_k$ among the equivalent subchannels. Moreover, this non-uniformity becomes more significant as the MIMO dimension increases. Therefore, spatiotemporal 2-D polar coding considering non-uniform subchannel reliabilities is essential for reliable transmission in practical MIMO systems.
\begin{figure}[!t]
	\centering
	\includegraphics[width=3.2in]{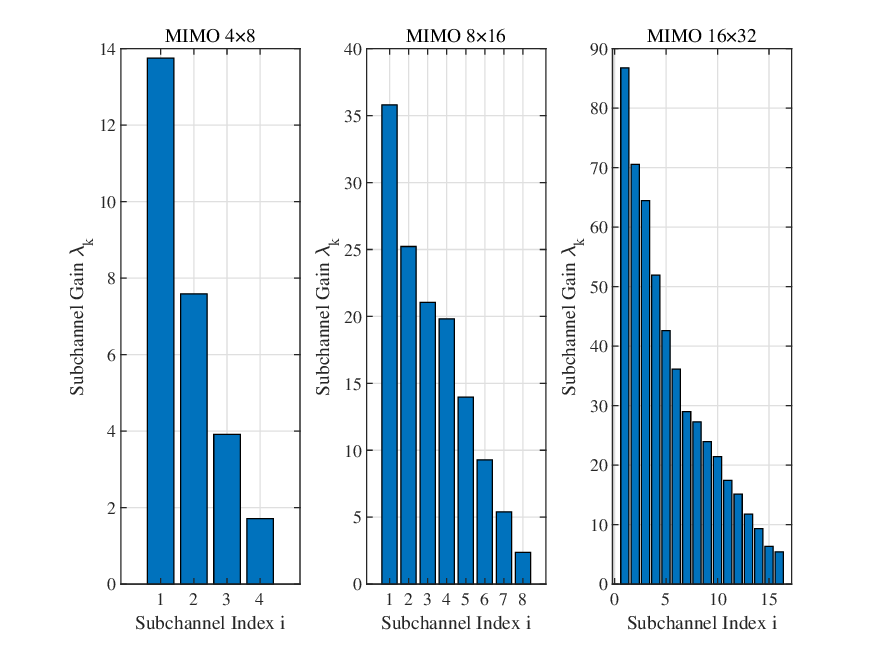}
	\caption{Distribution of subchannel gains for different MIMO configurations}
	\label{fig1}
\end{figure}

%Since the SVD transformation is unitary, it preserves mutual information, and thus the overall MIMO channel capacity satisfies
%\begin{equation}
%	\sum_{k=1}^{m} I(W_k) = I(W).
%\end{equation}

In the theory of spatiotemporal 2-D polar coding, channel polarization can be performed not only over virtual channels across different time slots, but also across different spatial channels. When considering spatiotemporal polarization over non-uniform spatial channels, the underlying subchannels exhibit distinct gains corresponding to the singular values ${\sqrt{\lambda_1},\ldots,\sqrt{\lambda_m}}$ obtained from the SVD of the MIMO channel. Building upon this observation, we propose a spatiotemporal 2-D polar coding framework for MIMO channels with non-uniform spatial gains, enabling effective polarization across both spatial and temporal dimensions.

\subsection{RCA-Based Spatiotemporal Polar-Code Construction}

Accurate frozen-bit selection is essential for fully utilizing the polarization effect, especially when the underlying parallel spatial streams exhibit markedly different reliabilities. In such non-uniform scenarios, the conventional Gaussian-approximation used for 1-D polar coding construction may lead to distorted reliability ordering, since it implicitly relies on a Gaussian-shaped LLR assumption. To obtain a more robust reliability metric under unequal spatial gains, we instead characterize the polarization evolution through an SNR-tracking viewpoint motivated by mutual-information duality, i.e., the reciprocal-channel principle. This line of approximation, commonly referred to as reciprocal channel approximation, evaluates reliability via capacity-domain mappings rather than enforcing a Gaussian LLR model, and is therefore better aligned with non-uniform subchannels.

The key idea is to relate an SNR value $\gamma$ to its ``reciprocal'' counterpart through the mutual information function $C(\gamma)$, and to propagate reliability on the polarization structure using this reciprocal mapping. Specifically, the reciprocal-channel mapping $\Psi(\gamma)$ is defined through the capacity complementarity $C(\Psi(\gamma)) = 1 - C(\gamma)$, and the resulting check-node update can be expressed via $\Psi(\cdot)$ while the variable-node update remains additive in $\gamma$. For numerical stability, SNR is expressed in the log domain $\xi \triangleq \ln \gamma$, and the mapping is implemented through $\Lambda(\xi)\triangleq \ln \Psi(e^\xi)$.

To reflect the unequal-gain nature of the MIMO channel in a spatiotemporal 2-D polar construction, we first diagonalize the MIMO channel via SVD. Assuming $S\leq L$, this yields $S$ parallel spatial streams, where the gain of the $k$-th stream is $\sqrt{\lambda_k}$ and its SNR is
\begin{equation}\label{gamma}
	\gamma_k = \lambda_k \cdot \frac{E_s}{N_0},
\end{equation}
where $E_s/N_0$ denotes the system-level SNR. This provides a direct quantification of spatial non-uniformity. Based on these stream-wise SNRs, we first form the spatial-domain log-reliability vector $\boldsymbol{\xi}_{\mathrm{sp}}^{(0)}=[\ln\gamma_1,\ln\gamma_2,\ldots,\ln\gamma_S]$. This $S$-dimensional spatial reliability profile is then replicated over the $T$ temporal positions to construct the equivalent 1-D initialization sequence $\boldsymbol{\xi}^{(0)}=[\boldsymbol{\xi}_{{\mathrm{sp}}}^{(0)},\boldsymbol{\xi}_{{\mathrm{sp}}}^{(0)},\ldots,\boldsymbol{\xi}_{{\mathrm{sp}}}^{(0)}]\in\mathbb{R}^{1\times(ST)}$, thereby embedding the spatial non-uniformity into the length-$N$ polarization process with $N=S \times T$.

The RCA-based reliability evolution and frozen-bit selection are then carried out in an equivalent 1-D polarization domain. This is justified by the established correspondence between the spatiotemporal 2-D polar transform and a length-$N$ 1-D polar transform with $N=S \times T$, under a bijection between the 2-D indices $(t,s)$ and the 1-D index $i$ \cite{li2025performanceanalysisspatiotemporal2d}. Accordingly, the RCA recursion is applied over the equivalent 1-D polarization tree under the conventional recursive reliability-propagation structure inherited from standard polar-code construction. For a pair of input reliabilities $\xi_0$ and $\xi_1$, the corresponding check-node and variable-node updates are given by \cite{2000On,Ochiai2023New}
\begin{equation}
	\hat{\xi}_{\mathrm{check}}
	=\Lambda\!\left(
	\max(\Lambda(\xi_{0}),\Lambda(\xi_{1}))
	+\ln\!\left(1+e^{-|\Lambda(\xi_{1})-\Lambda(\xi_{0})|}\right)
	\right),
\end{equation}
and
\begin{equation}
	\hat{\xi}_{\mathrm{var}}
	=\max(\xi_{0},\xi_{1})
	+\ln\!\left(1+e^{-|\xi_{0}-\xi_{1}|}\right),
\end{equation}
where $\Lambda(\cdot)$ is evaluated using the closed-form RCA approximation in \cite[Sec.~IV]{Ochiai2023New}. After recursively applying these updates to the full length-$N$ reliability sequence, the resulting reliability ordering and the associated frozen set are mapped back to the spatiotemporal grid, yielding the log-domain reliability set $\{\hat{\xi}_{t,s}\}$ for 2-D encoding/decoding. 

Importantly, since the initialization explicitly preserves the per-stream SNR differences (i.e., $\gamma_k$ varies with $\lambda_k$), the resulting reliability ordering better reflects the non-uniform MIMO characteristics, thereby mitigating the mis-ordering issues that may arise under single-parameter GA modeling and improving the transmission performance of spatiotemporal 2-D polar codes in realistic MIMO systems. In addition, this equivalent 1-D implementation preserves the low-complexity nature of the proposed scheme, since the RCA-based reliability evolution is carried out through closed-form updates without changing the underlying polarization recursion. Therefore, the proposed framework remains scalable to long blocklengths and large MIMO settings. Although the present work focuses on spatial-domain non-uniformity in MIMO systems, the underlying reliability-aware coding idea may also be extended to other fading-channel scenarios with heterogeneous parallel subchannels.

\subsection{NR-Compatible CSI Acquisition and Frozen-Set Alignment}\label{sec3:practical}

A practical requirement of the proposed spatiotemporal 2-D polar coding design is that the transmitter and receiver obtain a consistent reliability ordering for frozen-set selection, without dedicated signaling exchange. Specifically, both ends perform local channel estimation using standard reference signals already defined in 5G NR: the base station transmits downlink pilots and the user equipment transmits uplink sounding pilots (e.g., CSI-RS/SRS in NR). Based on their local channel estimates, both sides compute per-stream reliabilities and construct the frozen set independently. Since the frozen-set construction in this work depends only on the ordered singular-value spectrum through $\{\lambda_k\}$ and the induced per-stream SNRs in \eqref{gamma}, the two ends can derive aligned reliability orders from local measurements, eliminating the need to explicitly synchronize frozen-bit indices.

Moreover, the proposed mechanism naturally fits practical TDD operation. With reciprocity calibration, the effective uplink and downlink channels share approximately the same singular-value spectrum, which preserves the reliability ordering required by the polar-code construction.  

With imperfect CSI, each node only has an estimate $\widehat{\mathbf{H}}$, and the resulting spectrum $\{\widehat{\lambda}_k\}$ generally deviates from the true $\{\lambda_k\}$. Consequently, the frozen sets computed at the two ends may not be exactly identical. Nevertheless, the impact is limited in the practical regime considered in this work because (i) the construction depends primarily on the ordering of subchannel reliabilities rather than exact eigenvalues, and (ii) the proposed RCA-based spatiotemporal reliability propagation helps to mitigate mild mis-ordering caused by estimation noise. The resulting robustness is quantitatively verified by the simulations in Section~\ref{sec:result}.

\section{Simulation results and discussion}\label{sec:result}

In this section, simulation results are presented to evaluate the performance of the proposed spatiotemporal 2-D polar coding scheme over non-uniform MIMO channels. We consider a massive MIMO system, where the channel model is given in Section \ref{sec2}-A. For each channel realization, the simulated MIMO channel matrix is decomposed via SVD, and the resulting singular values are used to determine the corresponding per-stream SNRs. In the considered SVD-based MIMO transmission framework, transmitter-side CSI is assumed for spatial precoding. The code rate is set as 1/2, and the polar decoder is the successive cancellation (SC) decoder. Unless otherwise stated, BPSK modulation is employed, and the channel is estimated using a linear minimum mean square error (LMMSE) estimator with pilot length $L_p=2S$. The results demonstrate that accounting for spatial-domain reliability differences yields clear performance gains across various channel conditions, while maintaining compatibility with practical MIMO transceiver architectures.

\begin{figure}[!ht]
	\centering
	\subfigure[$S=4,L=8,T=32$]{
		\label{fig2_a}
		\includegraphics[width=3.2in]{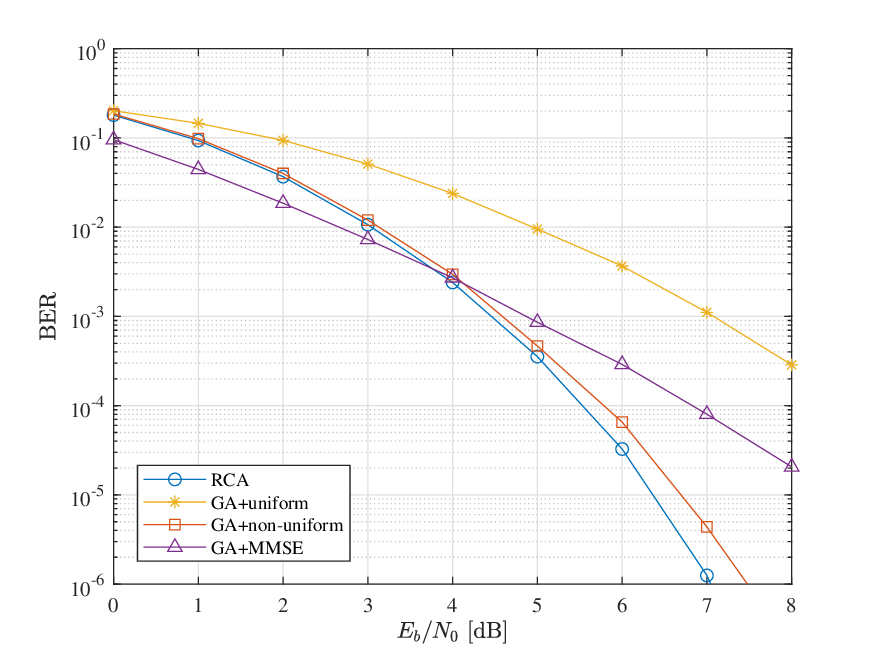}}
	\quad\\
	\subfigure[$S=8,L=16,T=32$]{
		\label{fig2_b}
		\includegraphics[width=3.2in]{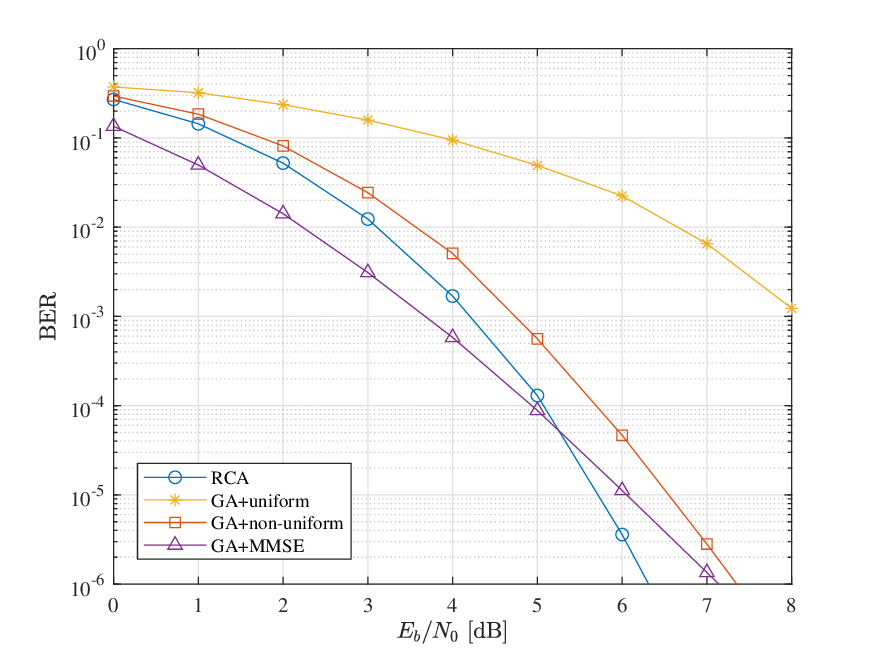}}
	\caption{The BER of spatiotemporal 2-D polar codes under different construction methods}
	\label{fig2}
\end{figure}

Figure \ref{fig2} shows the bit error rate (BER) performance of the proposed RCA-based spatiotemporal 2-D polar coding scheme together with three GA-based benchmarks: (i) SVD-based transmission with GA construction under a uniform-channel assumption, (ii) SVD-based transmission with GA construction that accounts for non-uniform MIMO channels, and (iii) GA construction combined with an MMSE receiver \cite{li2025performanceanalysisspatiotemporal2d}. Here, the first two SVD-based benchmarks are introduced to isolate the gain of the reliability-construction method itself under the same transceiver setting as the proposed scheme, whereas the GA+MMSE baseline is included mainly for comparison with the prior literature.

As observed, the uniform-channel GA construction suffers severe performance loss in practical non-uniform MIMO channels due to the violated identical-reliability assumption. This degradation further intensifies as $S$ increases (from 4 to 8), highlighting the necessity to explicitly model spatial reliability heterogeneity. The non-uniform GA partially alleviates this mismatch and performs reasonably well when the spatial dimension is small ($S=4$). However, its improvement remains limited, since the LLR statistics in heterogeneous MIMO channels deviate from the symmetric Gaussian model, and this limitation becomes more evident as non-uniformity strengthens. In contrast, the proposed RCA-based construction explicitly tracks spatial reliability variations without imposing Gaussian LLR assumptions, thereby consistently outperforming both GA-based counterparts under the same transceiver setting even with moderate spatial dimensions. Its advantage becomes increasingly evident as $S$ grows and is further amplified in the high-SNR regime, demonstrating strong robustness to heterogeneous MIMO conditions and more effective exploitation of spatiotemporal polarization.

The scheme in \cite{li2025performanceanalysisspatiotemporal2d} improves performance by leveraging the deterministic approximation behavior of MMSE detection, which effectively ``uniformizes'' the spatial streams. Note that the gain of this baseline is partially attributable to MMSE detection, rather than the construction method alone. Moreover, this benefit relies on sufficiently large antenna arrays and may be impractical in many realistic deployments. Nevertheless, although the GA with MMSE baseline may be competitive in the low-SNR region, the proposed RCA-based method exhibits a steeper BER improvement and becomes superior in the moderate-to-high SNR regime. This indicates that the proposed method remains competitive with the MMSE-based literature benchmark while offering a construction framework that is more directly tailored to non-uniform MIMO channels.

\begin{figure}[!ht]
	\centering
	\includegraphics[width=3.2in]{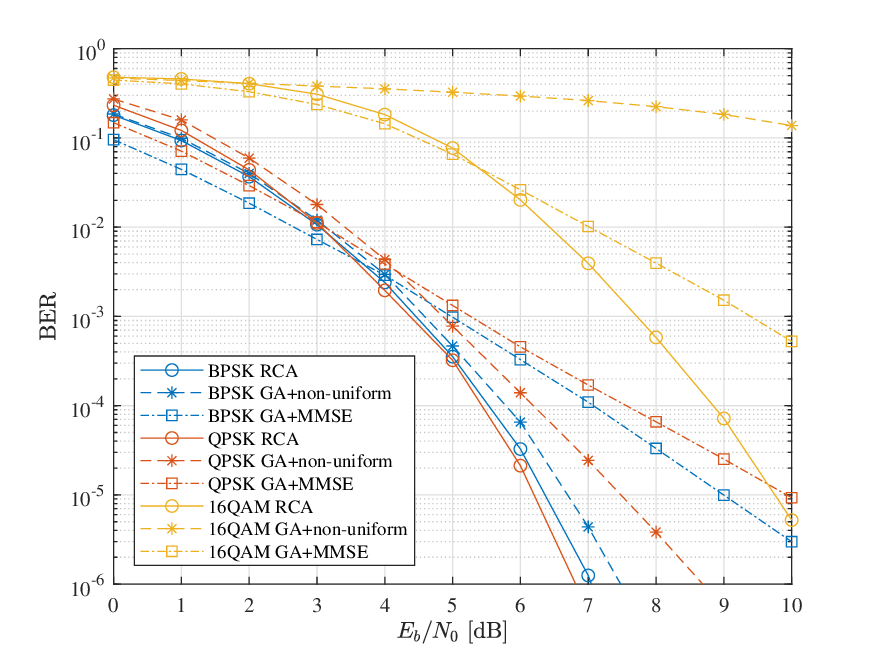}
	\caption{BER performance of the proposed 2-D coding under different modulation formats and construction methods ($S=4,L=8,T=32$)}
	\label{modulation}
\end{figure}

Figure \ref{modulation} shows the BER performance of the proposed spatiotemporal 2-D polar coding scheme under different modulation formats, including BPSK, QPSK, and 16QAM. As the modulation order increases, the BER of all schemes generally degrades. Nevertheless, the proposed RCA-based construction consistently achieves the best BER performance among the considered schemes under all tested modulation formats. This indicates that the proposed reliability-aware spatiotemporal 2-D polar coding framework remains effective and robust even under higher-order modulation settings. Meanwhile, it can also be observed that, as the modulation order increases, the performance advantage of RCA over the non-uniform GA-based construction becomes more pronounced, which is especially evident in the 16QAM case. This further confirms the advantage of RCA in handling the stronger heterogeneous reliability conditions jointly induced by higher-order modulation and non-uniform MIMO spatial channels.

\begin{figure}[!ht]
	\centering
	\includegraphics[width=3.2in]{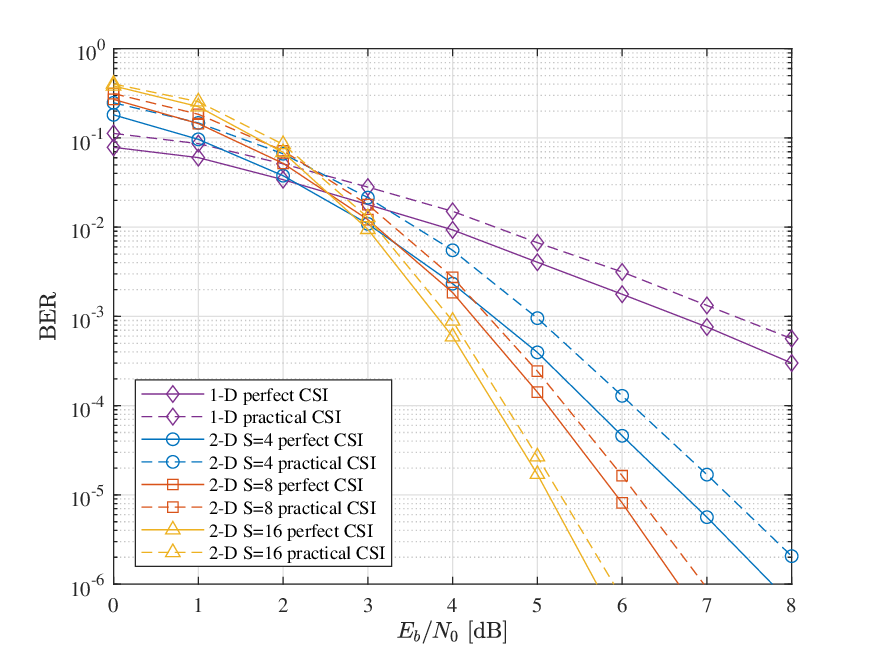}
	\caption{The BER of spatiotemporal 2-D polar codes under practical channel conditions ($T=32,L=2S$)}
	\label{fig3}
\end{figure}

Figure \ref{fig3} evaluates the BER performance in a realistic quasi-stationary MIMO channel with slowly varying large-scale fading. We consider two CSI cases: (i) perfect CSI, serving as an upper-bound reference, and (ii) imperfect CSI obtained from reference-signal-based LMMSE channel estimation, where the transmitter/receiver relies on the estimated channel for SVD operation, reliability evaluation and frozen-bit selection. Although estimation errors introduce a certain performance loss, the proposed RCA-based spatiotemporal 2-D polar coding remains highly robust because the relative reliability ordering across spatial streams is largely preserved within the quasi-stationary interval. Moreover, compared with conventional 1-D time-domain polar coding with the same temporal blocklength $T$, the proposed 2-D framework significantly improves reliability without increasing latency by additionally exploiting spatial-domain polarization. Increasing the spatial dimension $S$ further enhances performance, as larger spatial DoF strengthens the spatial polarization effect. These results confirm both practical robustness to CSI imperfections and scalable gains under realistic non-uniform MIMO conditions.

\section{Conclusion}\label{sec6}
In this paper, we propose a reliability-aware spatiotemporal 2-D polar coding framework tailored to practical MIMO channels with non-uniform spatial stream reliabilities. By embedding spatial SNR heterogeneity into the polarization and frozen-bit selection process through RCA, the proposed scheme effectively leverages spatial-domain reliability variations to improve transmission robustness. We further develop a practical implementation pathway compatible with 5G NR channel estimation mechanisms, enabling transceiver-assisted frozen-bit selection without additional signaling overhead. Simulation results demonstrate that the proposed RCA-based approach delivers consistent BER gains over conventional GA constructions designed for uniform reliability, remains robust under imperfect CSI, and achieves scalable reliability improvements as the spatial dimension increases. Overall, the proposed framework provides a practical, low-latency, and capacity-approaching solution for reliable MIMO transmission.

\ifCLASSOPTIONcaptionsoff
  \newpage
\fi

\balance
\bibliographystyle{IEEEtran}
\bibliography{ref}

@article{saad2019vision,
  title={A vision of {6G} wireless systems: Applications, trends, technologies, and open research problems},
  author={Saad, Walid and Bennis, Mehdi and Chen, Mingzhe},
  journal={IEEE Netw.},
  volume={34},
  number={3},
  pages={134--142},
  year={2020},
  month={May},
  publisher={IEEE}
}

@article{schulz2017latency,
  title={Latency critical {IoT} applications in {5G}: Perspective on the design of radio interface and network architecture},
  author={Schulz, Philipp and others},
  journal={IEEE Commun. Mag.},
  volume={55},
  number={2},
  pages={70--78},
  year={2017},
  month={Feb.},
  publisher={IEEE}
}

@article{you20236g,
  title={{6G} extreme connectivity via exploring spatiotemporal exchangeability},
  author={You, Xiaohu},
  journal={Sci. China Inf. Sci.},
  volume={66},
  number={3},
  pages={130306},
  year={2023},
  month={Mar.},
}

@ARTICLE{Ye2025ExplicitFBLMIMO,
  author={Ye, Feng and You, Xiaohu and Li, Jiamin and Ji, Chen and Zhang, Chuan},
  journal={IEEE Trans. Commun.},
  title={Explicit Performance Bound of Finite Blocklength Coded {MIMO}},
  note={early access, Aug. 1, 2025},
  year={2025},
  month={Aug.}
}

@inproceedings{you2022spatiotemporal,
  title={Spatiotemporal {2-D} channel coding for very low latency reliable {MIMO} transmission},
  author={You, Xiaohu and Zhang, Chuan and Sheng, Bin and Huang, Yongming and Ji, Chen and Shen, Yifei and Zhou, Wenyue and Liu, Jian},
  booktitle={Proc. IEEE Globecom Workshops (GC Wkshps)},
  pages={473--479},
  year={2022},
  month   = {Dec.}
}

@ARTICLE{li2025performanceanalysisspatiotemporal2d,
  author={Li, Yaqi and You, Xiaohu and Li, Jiamin and Ji, Chen and Sheng, Bin},
  journal={IEEE Transactions on Communications}, 
  title={Performance Analysis of Spatiotemporal {2-D} Polar Codes for Massive {MIMO} With {MMSE} Receivers}, 
  year={2026},
  volume={74},
  number={},
  pages={5284-5297},
}

@article{2000On,
  title={On the construction of some capacity-approaching coding schemes},
  author={ Chung, S. Y. },
  journal={Ph.D. dissertation, MIT},
  year={2000}
}

@article{Ochiai2023New,
  author={Ochiai, Hideki and Ikeya, Kosuke and Mitran, Patrick},
  journal={IEEE Trans. Commun.}, 
  title={A New Polar Code Design Based on Reciprocal Channel Approximation}, 
  year={2023},
  month={Feb.},
  volume={71},
  number={2},
  pages={631-643},
  publisher={Wiley Online Library}
}

@article{JUNG20111316,
  title   = {{REACT}: Rate Adaptation using Coherence Time in 802.11 {WLANs}},
  journal = {Computer Communications},
  volume  = {34},
  number  = {11},
  pages   = {1316--1327},
  year    = {2011},
  author  = {Jung, Hakyung and Kwon, Ted Taekyoung and Cho, Kideok and Choi, Yanghee}
}

@article{2000Capacity,
  title={Capacity of Multi-Antenna {Gaussian} Channels},
  author={ Telatar, Emre },
  journal={European Transactions on Telecommunications},
  volume={10},
  number={6},
  pages={585-595},
  year={1999},
}

@book{tse2005fundamentals,
  title={Fundamentals of wireless communication},
  author={Tse, David and Viswanath, Pramod},
  year={2005},
  publisher={Cambridge university press}
}

@ARTICLE{1192168,
  author={Gesbert, D. and Shafi, M. and Da-shan Shiu and Smith, P.J. and Naguib, A.},
  journal={IEEE Journal on Selected Areas in Communications}, 
  title={From theory to practice: an overview of {MIMO} space-time coded wireless systems}, 
  year={2003},
  volume={21},
  number={3},
  pages={281-302},
}

@ARTICLE{10255714,
  author={You, Xiaohu and Sheng, Bin and Huang, Yongming and Xu, Wei and Zhang, Chuan and Wang, Dongming and Zhu, Pengcheng and Ji, Chen},
  journal={IEEE Transactions on Communications}, 
  title={Closed-Form Approximation for Performance Bound of Finite Blocklength Massive {MIMO} Transmission}, 
  year={2023},
  volume={71},
  number={12},
  pages={6939-6951},
}

@ARTICLE{910580,
  author={Sae-Young Chung and Richardson, T.J. and Urbanke, R.L.},
  journal={IEEE Transactions on Information Theory}, 
  title={Analysis of sum-product decoding of low-density parity-check codes using a {Gaussian} approximation}, 
  year={2001},
  volume={47},
  number={2},
  pages={657-670},
}

@ARTICLE{6279525,
  author={Trifonov, Peter},
  journal={IEEE Transactions on Communications}, 
  title={Efficient Design and Decoding of Polar Codes}, 
  year={2012},
  volume={60},
  number={11},
  pages={3221-3227},
}

@ARTICLE{9220133,
  author={Ochiai, Hideki and Mitran, Patrick and Vincent Poor, H.},
  journal={IEEE Transactions on Communications}, 
  title={Capacity-Approaching Polar Codes With Long Codewords and Successive Cancellation Decoding Based on Improved {Gaussian} Approximation}, 
  year={2021},
  volume={69},
  number={1},
  pages={31-43},
}

\end{document}